\documentclass[aps,pra,twocolumn,superscriptaddress]{revtex4-2}
\usepackage[T1]{fontenc}
\usepackage[utf8]{inputenc}
\usepackage{soul}
\usepackage{textcomp}
\usepackage[english]{babel}
\usepackage{graphicx}
\usepackage{dcolumn}
\usepackage{bm}
\usepackage{bm}
\usepackage{amsmath}
\usepackage{verbatim}     
\usepackage[dvipsnames]{xcolor}       
\usepackage[normalem]{ulem}   
\usepackage{bbold}
\usepackage{mathrsfs}
\usepackage[mathscr]{eucal}
\usepackage{ulem}
\usepackage{physics}
\newcommand{\beq}{\begin{equation}}
\newcommand{\eeq}{\end{equation}}
\newcommand{\bei}{\begin{itemize}}			
\newcommand{\eei}{\end{itemize}}			

\newcommand{\vect}[1]{{\mathbf #1}}
 
\usepackage{hyperref}
\hypersetup{
   pdfpagemode=None,
   pdfstartpage=1,
   pdfmenubar=true,
   pdftoolbar=true,
   colorlinks = true,
   linkcolor=blue,
   citecolor=blue,
   urlcolor=blue,
   bookmarksopen=false
 }

\begin{document}
\title{Large-scale free-space photonic circuits in two dimensions}
\author{Maria Gorizia Ammendola}
\thanks{MGA and FDC contributed equally to this work.}
\affiliation{Dipartimento di Fisica, Universit\`{a} degli Studi di Napoli Federico II, Complesso Universitario di Monte Sant'Angelo, Via Cintia, 80126 Napoli, Italy}
\affiliation{Scuola Superiore Meridionale, Via Mezzocannone, 4, 80138 Napoli, Italy}
\author{Francesco Di Colandrea}
\email{francesco.dicolandrea@unina.it}
\affiliation{Dipartimento di Fisica, Universit\`{a} degli Studi di Napoli Federico II, Complesso Universitario di Monte Sant'Angelo, Via Cintia, 80126 Napoli, Italy}
\affiliation{Nexus for Quantum Technologies, University of Ottawa, K1N 5N6, Ottawa, ON, Canada}
\author{Lorenzo Marrucci}
\affiliation{Dipartimento di Fisica, Universit\`{a} degli Studi di Napoli Federico II, Complesso Universitario di Monte Sant'Angelo, Via Cintia, 80126 Napoli, Italy}
\affiliation{CNR-ISASI, Institute of Applied Science and Intelligent Systems, Via Campi Flegrei 34, 80078 Pozzuoli (NA), Italy}
\author{Filippo Cardano}\email{filippo.cardano2@unina.it}
\affiliation{Dipartimento di Fisica, Universit\`{a} degli Studi di Napoli Federico II, Complesso Universitario di Monte Sant'Angelo, Via Cintia, 80126 Napoli, Italy}

\begin{abstract}
\noindent
Photonic circuits, engineered to couple optical modes according to a specific map, serve as processors for classical and quantum light. The number of components typically scales with that of processed modes, thus correlating system size, circuit complexity, and optical losses. Here we present a photonic-circuit technology implementing large-scale unitary maps in free space, coupling a single input to hundreds of output modes in a two-dimensional compact layout. The map corresponds to a quantum walk of structured photons, realized through light propagation in three liquid-crystal metasurfaces, having their optic axes artificially patterned. Theoretically, the walk length and the number of connected modes can be arbitrary, while keeping losses constant. The patterns can be designed to replicate multiple unitary maps. We also discuss limited reconfigurability by adjusting the overall birefringence and the relative displacement of the optical elements. These results lay the basis for the design of low-loss non-integrated photonic circuits, primarily for manipulating multi-photon states in quantum regimes.
\end{abstract}

\maketitle

\section{Introduction}
Optical degrees of freedom, such as those associated with spatial, spectro-temporal, or polarization features of the optical field, serve as a convenient resource for encoding information. The abundance of tools for their accurate manipulation established photonics as a versatile platform for both classical and quantum information processing tasks.

Currently, a variety of platforms have been demonstrated that can realize different operations on optical modes \cite{Bogaerts2020}, including vector-matrix multiplication~\cite{Nikkhah2024c}, nonlinear maps~\cite{Kirsch2021b}, and unitary transformations~\cite{Hoch20220}. Optical processors based on linear circuits provide key applications in optical computing ~\cite{OBrien2007,McMahon2023} and are emerging as building blocks of future optical neural networks and AI systems~\cite{Wetzstein2020}. 

When used as optical simulators, the processed optical modes encode the degrees of freedom of a target system (typically, lattice models describing electronic systems in condensed matter), and the overall transformation maps to a unitary temporal evolution operator. By monitoring the system output one can observe directly optical analogues of classical or quantum dynamics~\cite{Haldane2008,Walther2012,Szameit19}.

Quantum light may be injected at the input ports of these systems, yielding output states strongly affected by the quantum interference of two (or more) photons~\cite{Hiekkamaki2021, Bromberg2022, Defienne2023, Makowski, Goel2240}. The complexity associated with multi-particle interference phenomena underlies Boson sampling problems, extremely popular in the recent past as they provided the playground for the first instances of quantum advantage~\cite{Zhong2020,Arrazola2021}.

Optical platforms like those mentioned above, performing a variety of tasks, are often referred to as photonic circuits. This classification considers their analogy with other circuits where the information carriers, like electric signals, are routed to distinct channels and processed. In integrated systems, this analogy is straightforward, as optical signals (both as macroscopic wave-packets or single photons) are spatially localized (like electrical currents), travel along distinct waveguides, and are manipulated through integrated beam splitters and phase shifters~\cite{Wang2020}.

Optical modes building a photonic circuit may not correspond to separated paths for traveling light, but may correspond to co-propagating modes that are orthogonal because of alternative degrees of freedom like spectro-temporal ones or those associated with transverse spatial modes, like those carrying orbital angular momentum~\cite{Flamini2018rome,Slussarenko2019}. 

In the first case, trains of pulses associated with non-overlapping time-bins are conveniently manipulated via propagation into fibers, paths of different lengths, and birefringent materials, with a variety 
of applications such as quantum information processing~\cite{fred_ultrafast_2}, quantum walks \cite{Fenwick2024, Wang2024b}, quantum computing~\cite{Arrazola2021}, and quantum communication~\cite{Bouchard2022}. 

In the second case, the manipulation of co-propagating transverse modes of structured light via propagation through multi-mode fibers, complex diffractive elements, or multi-plane light converters~\cite{Matthes2019, Cristiani22,Piccardo21,Kupianskyi230} has been successfully demonstrated in the recent years. While these alternative circuits have not yet reached the technological maturity of integrated solutions, they offer advantages in terms of the number of addressable modes, reconfigurability, and the alternative detection schemes of quantum light by using camera-like sensors~\cite{Makowski}.

Within this context, photonic circuits based on liquid-crystal metasurfaces (LCMSs) have been recently introduced for the realization of quantum walks (QWs). A LCMS is an ultra-thin, transmissive plate made of a micrometric layer of LC molecules, with their local orientation being artificially patterned at the micrometric scale. Essentially, they act as standard waveplates for polarization manipulation, but with a spatially varying optic-axis orientation~\cite{Rubano2019}. When exhibiting periodic patterns, LCMSs couple circularly polarized modes of light that carry quantized transverse momentum~\cite{DErrico2020}.

The original scheme for the realization of QWs with this platform required a long sequence of periodic LCMSs, coupling modes arranged both in 1D~\cite{DErrico2020b} and 2D~\cite{DErrico2020} grids. In the 1D case, a technique has been recently demonstrated that allows compressing the entire transformation into only three metasurfaces~\cite{DiColandrea2023}. This result is independent of the walk length and the number of involved modes, which is strictly related to the size of the implemented unitary matrix, thus dramatically reducing optical losses. 

To fully exploit the two-dimensional nature of transverse modes, it would be highly desirable to implement this concept with modes arranged in a 2D grid. However, a direct application of the same ``compression'' method used in the 1D case presents an unexpected challenge: the existence of multiple solutions to the problem typically results in the formation of several discontinuity lines in the 2D LCMS pattern. These discontinuous patterns are inherently unstable in LC media, rendering LCMS devices with such patterns generally nonfunctional. In this work, we propose and experimentally validate a scheme that effectively addresses this issue by consistently identifying quasi-continuous solutions for the 2D patterns. This approach eliminates all discontinuities except for localized point singularities, which are topological in nature and therefore unavoidable, but have negligible optical effect. We then report instances of unitary transformations that are equivalent to 2D QWs up to $20$ time steps, mapping localized inputs to superpositions of up to $800$ modes arranged in a square lattice. The same amount of modes would have required hundreds of time steps in the 1D case, leading to beams with much higher transverse momentum, which inherently suffer faster diffraction and need larger camera sensors to be detected.
\begin{figure}
    \includegraphics[width=\linewidth]{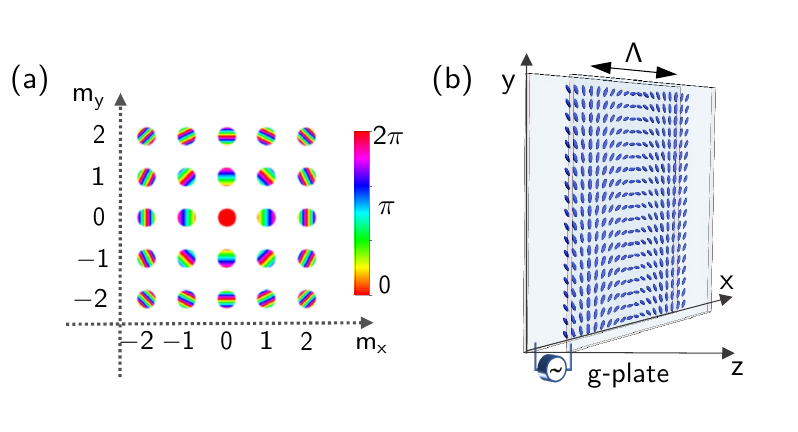}
    \caption{\textbf{QWs in the space of light transverse momentum.}
    (a)~Photonic modes implementing the position states on the lattice. For each mode carrying $m_x$ and $m_y$ units of transverse momentum $\Delta k_{\perp}$ in the $x$ and $y$ directions, respectively, we plot the linear phase profile in the transverse $xy$ plane.
    (b)~LC pattern of a $g$-plate. The local molecular director forms an angle $\theta$ with the $x$ axis. In a $g$-plate, we have $\theta(x)=\pi x/\Lambda$, with $\Lambda$ being the spatial period. The birefringence is uniform and electrically tunable by applying a voltage to the cell~\cite{Piccirillo2010}.}
    \label{fig:apparatus}
\end{figure}

\section{Materials and Methods}
\subsection{QWs in the light transverse momentum space via LCMSs}
\begin{figure*}[t]
    \includegraphics[width=0.95\linewidth]{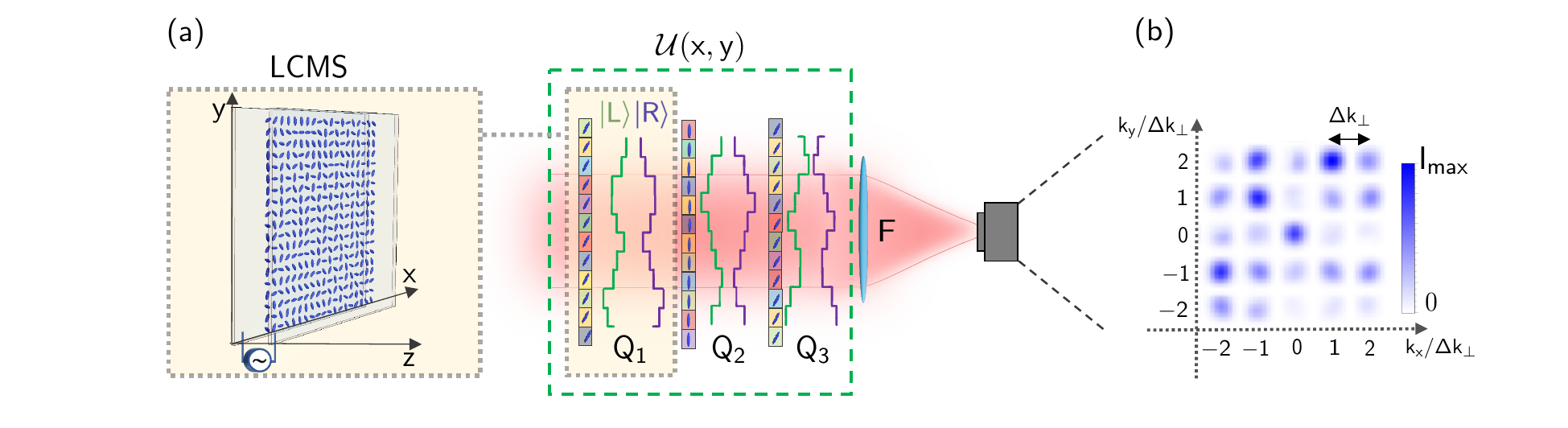}
    \caption{\textbf{Large-scale mode mixing via LCMSs.} (a)~Three LCMSs ($Q_1, Q_2, Q_3$) implement the optical transformation corresponding to the desired multi-mode mixing $\mathcal{U}$. The inset illustrates a LCMS with its LC optic-axis pattern. Squares of different colors reflect different values of ${\theta (x,y)}$, while the birefringence $\delta$, i.e., the orientation with respect to the propagation axis $z$, is homogeneous. 
    Off-diagonal elements of the LCMS Jones matrix flip the polarization handedness and add a space-dependent conjugate phase modulation on orthogonal circular polarization components $\ket{L}$ and $\ket{R}$. 
    (b)~The mode sorting is realized in the focal plane of a lens (F), where modes appear as a 2D array of Gaussian beams separated by $\Delta k_{\perp}$. Each spot is a superposition of the polarization (coin) states $\{\ket{L}, \ket{R}\}$.}
    \label{fig:apparatus2}
\end{figure*}
QWs represent a convenient framework for building unitary transformations to be directly implemented in a photonic circuit. These are prototypical discrete-time dynamics of a particle (walker) on a lattice, whose motion is conditioned by a spin-like internal degree of freedom (coin). In a 2D configuration, position states $\ket{m_x,m_y}$ ($m_x, m_y$ are integers), associated with lattice sites and spanning the Hilbert space $\mathcal{H}_\ell$, are combined with coin states $(\ket{0},\ket{1})$ spanning the space $\mathcal{H}_s$ to form the circuit modes. We consider two-level coin systems and assume that the input state is prepared in a single mode, that is a localized walker ${\ket{\psi_0}=\ket{m_x=0,m_y=0}\otimes\ket{\phi_0}}$, where $\ket{\phi_0}$ is the input coin state. After $t$ steps, the system is mapped into a linear superposition of multiple modes, whose number scales linearly with step number: 
\begin{align}\label{qwprocess}
\begin{split}
    \ket{\psi_t}&=U_0^t \ket{\psi_0}=\\
    &=\sum_{m_x,m_y} \sum_{j \in \lbrace 0,1\rbrace}c_{m_x,m_y,j}\ket{m_x,m_y}\otimes \ket{j}.
\end{split}
\end{align}
Here, $U_0$ is the single-step evolution operator. We assume this operator to be identical at each step, though this condition can be relaxed to obtain more general transformations associated with time-dependent QWs.

In this paper, we focus on the QWs introduced in Ref.~\cite{DErrico2020}, where the single-step evolution operator is
\begin{equation}
U_0(\alpha)=T_y(\alpha)T_x(\alpha)W.
\end{equation}
Here, $W$ is the coin rotation operator, reading
\begin{equation}
W=\frac{1}{\sqrt{2}}
\begin{pmatrix}
    1 && i\\
    i && 1
\end{pmatrix},
\end{equation}
and
\begin{equation}\label{eqn:Tx}
T_x(\alpha)=\begin{pmatrix}
\cos(\alpha/2) && i\sin(\alpha/2)\hat{t}_x\\
i\sin(\alpha/2)\hat{t}^\dagger_x && \cos(\alpha/2)
\end{pmatrix}
\end{equation}
is the translation operator along $m_x$, with ${\hat{t}_x\ket{m_x,m_y}=\ket{m_x-1,m_y}}$. A similar expression holds for $T_y(\alpha)$. The parameter $\alpha$ tunes the hopping amplitudes between neighboring sites. We specifically set $\alpha=\pi/2$. We anticipate here that our method, outlined in the next section, applies to any form of translation and coin operators, as long as their parameters are independent of the walker position, i.e., there is translation invariance. In the Supplementary Material, we present numerical simulations for a different fully-balanced 2D QW studied elsewhere~\cite{Esposito2022}.

Our photonic implementation of the QW states defined in Eq.~\eqref{qwprocess} employs optical modes having the following expression:
\begin{equation}\label{opticalmodes}
    \ket{m_x,m_y,j}= A(x,y,z) e^{i k_z z} e^{i (m_x x+m_y y) \Delta k_{\perp}}\ket{j},
\end{equation}
where $A(x,y,z)$ is a Gaussian envelope with a beam waist $w_0$, $k_z$ is the wavevector $z$ component, $ \Delta k_{\perp}$ is a unit of transverse momentum, and $\ket{j}$ is a left/right circular polarization state $\ket{L}$/$\ket{R}$, respectively (see Fig.~\ref{fig:apparatus}(a)). To have a negligible cross-talk between these modes, their waist radius must be greater than $2\pi/\Delta k_\perp$ \cite{DErrico2020}.
The most straightforward way to engineer the QW dynamics with these modes is by cascading a sequence of polarization gratings having ${\Lambda=2\pi/\Delta k_{\perp}}$ as their spatial period. Intuitively, these give photons a transverse momentum kick equal to $\pm \Delta k_{\perp}$, depending on the polarization being left or right circular, respectively, thus implementing the QW shift operator. This is the key idea at the basis of the first experiment demonstrating QWs with such transverse modes, with polarization gratings realized in terms of LCMSs termed \emph{g}-plates \cite{DErrico2020}
(see Fig.~\ref{fig:apparatus}(b)).

As anticipated, LCMSs consist of a micrometric nematic LC layer sandwiched between two glass plates, whose internal sides are coated with a transparent conductive material to enable the application of electric fields.
Such devices can be modeled as standard waveplates with an inhomogeneous optic-axis orientation. In the circular polarization basis, their Jones matrix reads
\beq\label{jones}
    Q_\delta(\theta)= 
    \begin{pmatrix}
    \cos(\delta/2) &&i\sin(\delta/2) e^{-2 i \theta(x,y)}\\
    i\sin(\delta/2) e^{2 i \theta(x,y)} && \cos(\delta/2)
    \end{pmatrix}.
\eeq
Here, $\delta$ is the optical birefringence parameter determined by the out-of-plane tilt angle of LC molecules, 
and $\theta(x,y)$ is the optic-axis in-plane orientation with respect to the reference $x$ axis. 
Patterns of LC orientation are imprinted via a photoalignment technique~\cite{Rubano2019}. The birefringence parameter $\delta$ is uniform across the cell but can be adjusted by tuning an electric field~\cite{Piccirillo2010}. Diagonal elements of the LCMS matrix (proportional to $\cos{(\delta/2)}$) leave part of the beam unaltered. Off-diagonal elements (proportional to $\sin{(\delta/2)}$) flip the polarization handedness and add a space-dependent geometric phase (equal to $2\theta$ and opposite for orthogonal circular polarizations), as pictorially shown in Fig.~\ref{fig:apparatus2}(a).
The action of a \emph{g}-plate, where ${\theta(x)=\pi x/\Lambda}$, is equivalent to the translation operator of Eq.~\eqref{eqn:Tx}, with $\alpha=\delta$.
Using classical light, 1D QWs up to 14 steps~\cite{DErrico2020b, DErrico2021} and 2D QWs up to 5 steps~\cite{DErrico2020} have been realized via the action of several cascaded \emph{g}-plates. Using a two-photon input state, 3 steps of a 2D QW have been reported, with the walk length limited by the number of available single-photon detectors and optical losses~\cite{Esposito2022}. The latter indeed represents a key limiting factor in multi-photon experiments. The number of devices (or the circuit depth in integrated architectures) scales linearly with the number of steps, therefore losses increase exponentially and severely limit the possibility of implementing large-scale evolutions in a genuinely quantum regime.

\subsection{Large-scale mode mixing via three LCMSs}
\label{sec:minimal}
In typical experiments using LCMSs to realize photonic circuits for QWs,  diffraction between consecutive devices is avoided. As such, the action of a long sequence of LCMSs is described by the product of the Jones matrices of individual LCMSs, each featuring the form of Eq.~\eqref{jones}. The resulting matrix is thus the Jones matrix associated with the entire system, having spatial frequencies that increase with the number of steps to be realized. 

The entire sequence can be replaced by a shorter chain of LCMSs. It is well known indeed that an arbitrary polarization transformation can be realized via a minimal sequence of three waveplates~\cite{Simon1990,Sit2017}. A common choice is a half-wave plate sandwiched between two quarter-wave plates, that is $Q_{\pi/2}(\theta_3)Q_{\pi}(\theta_2)Q_{\pi/2}(\theta_1)$ (see Eq.~\eqref{jones}).

By leveraging the possibility of patterning the optic axis of LCMSs, we build on this idea to realize arbitrary space-dependent transformations, modeled in terms of inhomogeneous Jones matrices~\cite{DiColandrea2023}.  At each transverse position, the target unitary $\mathcal{U}(x,y)$ can be implemented by three LCMSs, whose LCs are locally oriented to satisfy the equation (see Fig.~\ref{fig:apparatus2}(a)):
\beq\label{eqn:solve}
   \mathcal{U}(x,y)=Q_{\pi/2}(\theta_3(x,y))Q_{\pi}(\theta_2(x,y))Q_{\pi/2}(\theta_1(x,y)),
\eeq 

\begin{figure}[h!]
\includegraphics[width=\linewidth]{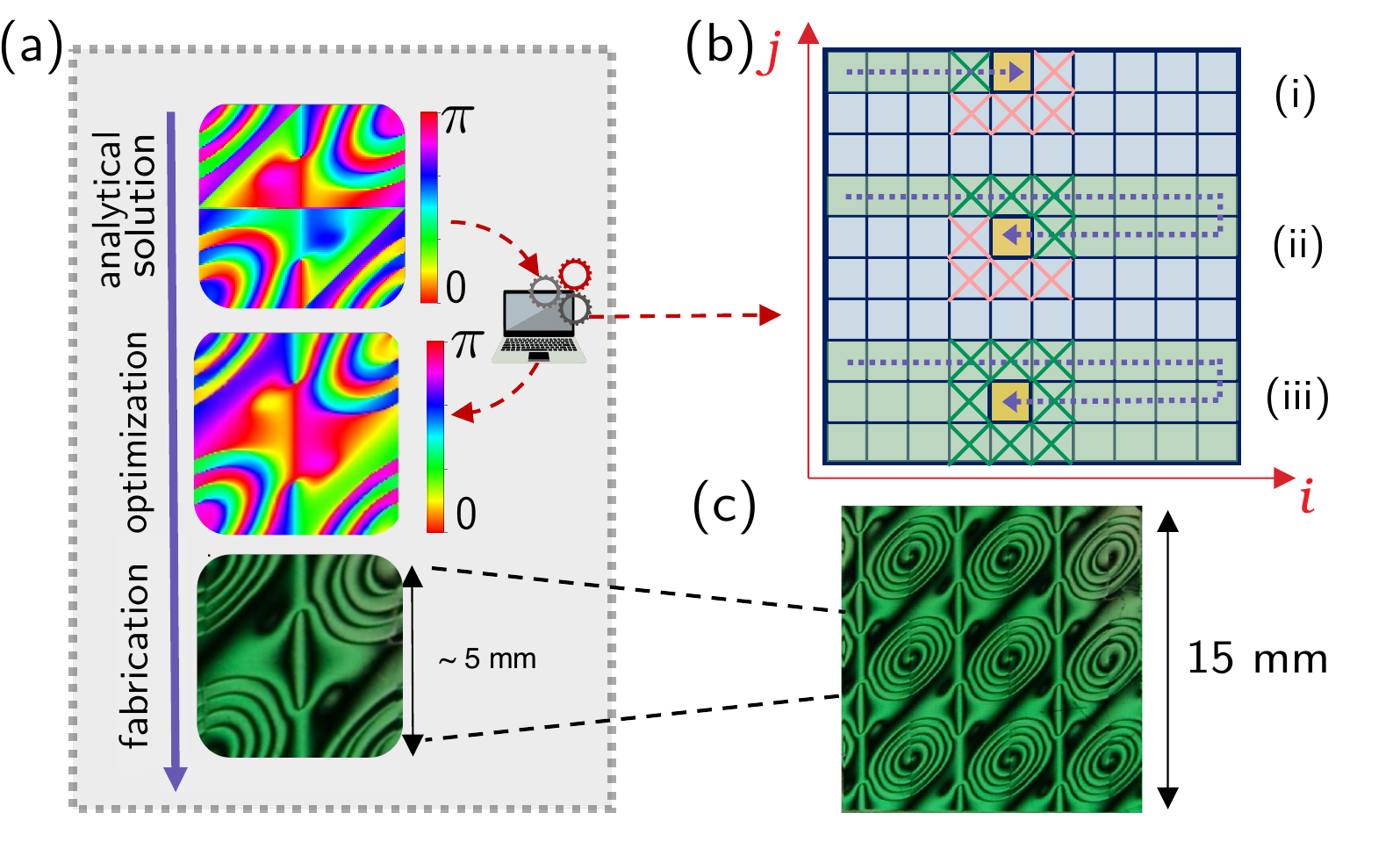}
    \caption{\textbf{Numerical optimization to retrieve 2D continuous LCMS optic-axis modulations.}
    (a)~Values of $\theta(x,y)$ obtained from a single set of analytical solutions of Eq.~\eqref{eqn:solve} are typically discontinuous. A numerical routine is devised to match different solutions at each transverse position to enforce continuity.
    The resulting pattern displays isolated vortices.
    (b)~Different scenarios (i)-(ii)-(iii) are illustrated for the optimization routine, depending on the current position on the plate $\vect{r_{ij}}$ (yellow square). The violet arrow path contains discrete positions where the optimization algorithm has already been executed. The green crosses mark neighboring elements $\vect{r_{n}}$ where a continuous modulation has already been found, and are therefore involved in the optimization of the metric $d_{ij}$ (see text). Neighboring elements where the algorithm has not been executed yet are marked by red crosses. (c)~Full pattern of one of the LCMSs designed to implement a 10-step QW ($3 \Lambda \times 3 \Lambda$ square, with $\Lambda=5$ mm), imaged between crossed polarizers to reveal the LC's in-plane orientation.}
    \label{fig:search}
\end{figure}

To achieve this goal, we first compute the overall Jones matrix $\mathcal U$ associated with the entire walk, and then solve Eq.~\eqref{eqn:solve} in terms of $\theta_1,\theta_2,\theta_3$ at each transverse position. As in our original formulation in the context of 1D QWs~\cite{DiColandrea2023}, these equations admit two distinct sets of analytical solutions. Differently from the former case, however, modulations provided by these analytical solutions yield discontinuous LC patterns, exhibiting several extended lines of disclinations (see Fig.~\ref{fig:search}(a), top). These 2D discontinuities correspond to LC domain walls, which are inherently unstable due to LC elasticity, so that the corresponding LCMSs would show strongly non-ideal optical behavior.

\begin{figure*}[t]
    \includegraphics[width=\linewidth]{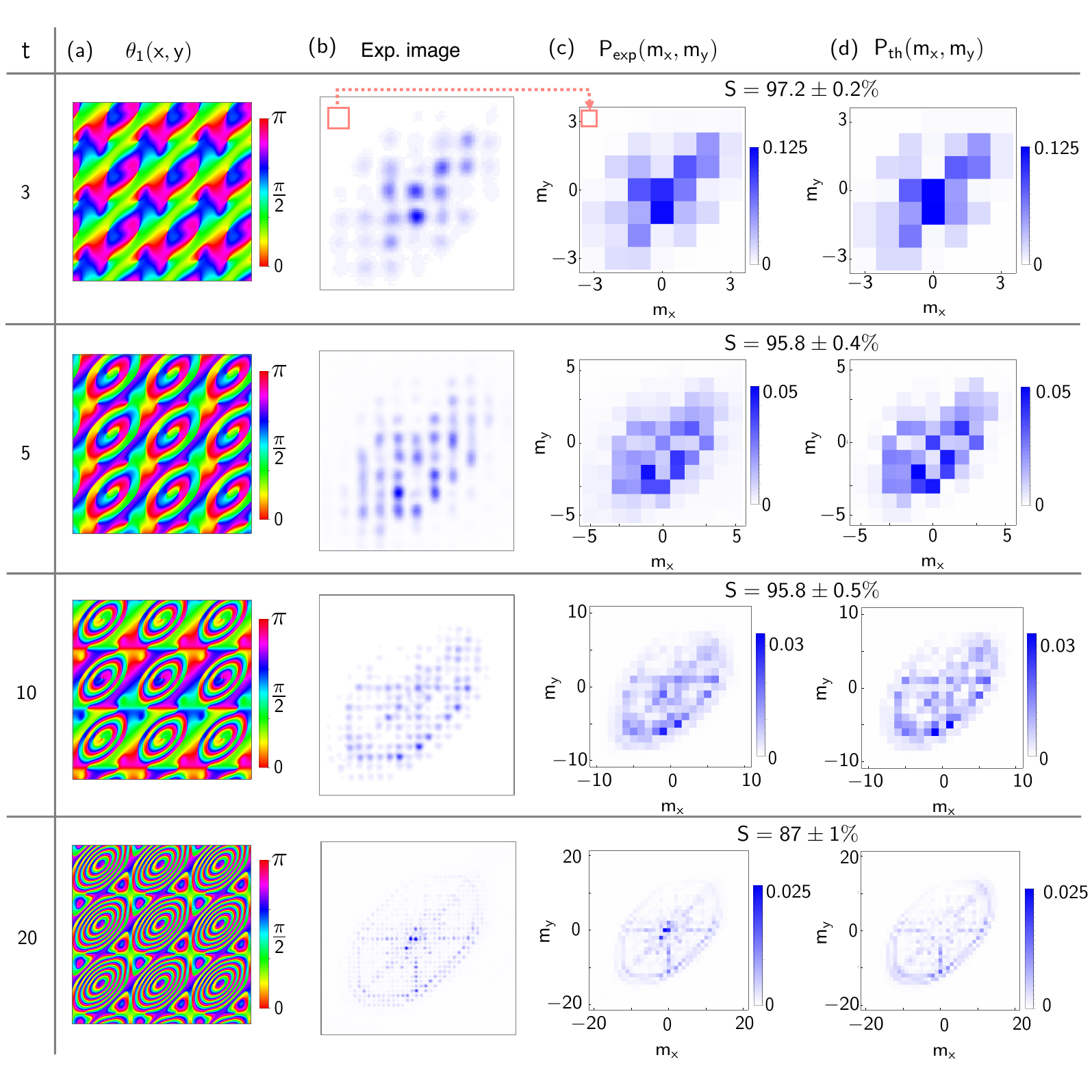}
    \caption{\textbf{2D QWs via spin-orbit photonics.}
   (a)~Optic-axis modulation of the first metasurface ($\theta_1(x,y)$) employed for the simulation of the 2D QW. (b)~Experimental images obtained for a $\ket{R}$-polarized input state, from which the walker probability distribution $P_{\text{exp}}(m_x,m_y)$ is extracted (c), and compared with the theoretical prediction $P_{\text{th}}(m_x,m_y)$ (d). For each realization, we report the value of the similarity, computed as the average of four independent measurements. 
   Rows refer to $3$, $5$, $10$, and $20$ time steps ($t$), respectively.}
\label{fig:exp_data}
\end{figure*}

To tackle this issue, we developed an optimization routine enabling us to enforce continuous patterns for LC angles (see Fig.~\ref{fig:search}(a), center). This procedure tolerates singularities for the LC orientation, appearing as vortices carrying a topological charge, clearly visible in the central panel of Fig.~\ref{fig:search}(a) and in the other patterns presented in Fig.~\ref{fig:exp_data}. The vortex charge quantifies the rotation of LC molecules (modulo $\pi$) when following a closed trajectory around the singular point. Interestingly, the elementary charges are always $\pm 1/2$, which are stable. Any higher order vortex $|N|>1/2$ would split in $N$ elementary vortices in a real device~\cite{stewart2004book}. 

The optimization proceeds as follows. To solve Eq.~\eqref{eqn:solve}, we decompose the two sides in terms of the generators of SU(2):
\begin{equation}\label{process}
     \sum_{i=0}^3c_{i}(x,y)\sigma_{i}=\sum_{i=0}^3\ell_{i}(x,y)\sigma_{i},
 \end{equation}
 where $\sigma_0$ is the identity matrix and $\sigma_1$, $\sigma_2$, and $\sigma_3$ are the  Pauli matrices.
By matching terms from both sides, one can determine the optic-axis modulations for the three LCMSs: $\theta_{\alpha}(x,y)$, $\alpha \in \{1,2,3\}$. 
The analytical solution of Eq.~\eqref{process} yielding the sets of patterns for a fixed number of steps is provided in Appendix~\ref{sec:methods2}.

To ensure continuity, the algorithm minimizes a metric that quantifies the difference between the optic-axis orientations at neighboring positions:
\begin{equation}
    d_{ij}=\sum_{n=1}^{N_{ij}} \bigg[ \sum_{\alpha =1}^3 \left(\theta_{\alpha}(\vect{r_{ij}}) -  \theta_{\alpha}( \vect{r_{n}})\right)^2\bigg].
    \label{eqn:metric}
\end{equation} The latter provides a measure of the distance between the orientation of the optic axis of the three LCMSs at the current transverse position $\vect{r_{ij}}$ and its $N_{ij}$ neighboring elements $\vect{r_{n}}$ (chosen within a tunable range), where a possible modulation has already been found.
The working principle of the algorithm is illustrated in Fig.~\ref{fig:search}(b).

As expected, the complexity of such modulations increases with the complexity of the simulated process. This is also evident in the patterns shown in Fig.~\ref{fig:exp_data}(a), where we plot the optic-axis modulation of the first LCMS ($\theta_1(x, y)$) employed for the simulation of $3$, $5$, $10$, and $20$ steps of the QW protocol described above. The minimal transformation naturally preserves the spatial periodicity $\Lambda$ characterizing the original cascaded scheme.
The plotted modulations are relative to a $3\Lambda \times 3\Lambda$ square, with $\Lambda = 5$~mm in the experiment. 

The final stage of a mode-mixing experiment consists of the mode sorting and detection stage. The modes of Eq.~\eqref{opticalmodes} can be spatially resolved on a CCD camera placed in a focal plane of a lens, implementing an optical Fourier Transform (see Fig.~\ref{fig:apparatus2}(b)). As discussed above, these modes have negligible overlap as long as $w_0\geq\Lambda$~\cite{DErrico2020}, where $w_0$ is the beam waist. A complete description of the experimental setup is provided in Appendix~\ref{sec:methods1}. 

\section{Results}
Representative experimental images for a $\ket{R}$-polarized localized input after ${3,5,10,}$ and 20 steps are shown in Fig.~\ref{fig:exp_data}(b), from which the QW probability distributions can be extracted (see Fig.~\ref{fig:exp_data}(c)). Each light spot is associated with a walker site, with probability given by the integrated and normalized light intensity within that spot. The output modes distribution is directly related to the unitary map, in this case, our specific QW protocol. The specific orientation of the walker distribution reflects the structure of the QW protocol, that misses a coin rotation between consecutive translations along the $x$ and $y$ directions~\cite{DErrico2020}. When adding such additional operation, the walker symmetrically spreads across the entire lattice (see Ref.~\cite{Esposito2022} and the Supplementary Material).
The procedure to extract the walker probability distribution is outlined in Appendix~\ref{sec:methods3}. Figure~\ref{fig:exp_data}(d) shows the corresponding theoretical probability distributions. The agreement between the theoretical predictions and the experimental observations is quantified in terms of the similarity
\begin{equation}\label{similarity}
S=\left(\sum_{m_x,m_y}\sqrt{P_\text{{exp}}(m_x,m_y) P_\text{{th}}(m_x,m_y)}\right)^2,  
\end{equation}
where $P_\text{{exp}}$ and $P_\text{{th}}$ are the normalized experimental and theoretical probability distributions, respectively.
A good agreement with the theory is observed in all our experiments, with similarity always above $87\%$. The uncertainties are computed as the standard deviation over $4$ independent measurements. The decrease in similarity observed with the increase in the number of modes is ascribed to the increasing complexity of the LCMS patterns when targeting longer evolutions.   Experimental results obtained with different input coin states are reported in the Supplementary Material. The circuit efficiency here does not decrease with the number of steps, as the number of optical components stays constant, with an average recorded total transmission of $69 \pm 1 \%$ from the input to the output of the QW platform. In a traditional setup, after 20 steps a similar overall transmittance would be only possible with a single-step efficiency greater than 98\%.
 
\begin{figure}[t]
\centering
\includegraphics[width=\linewidth]{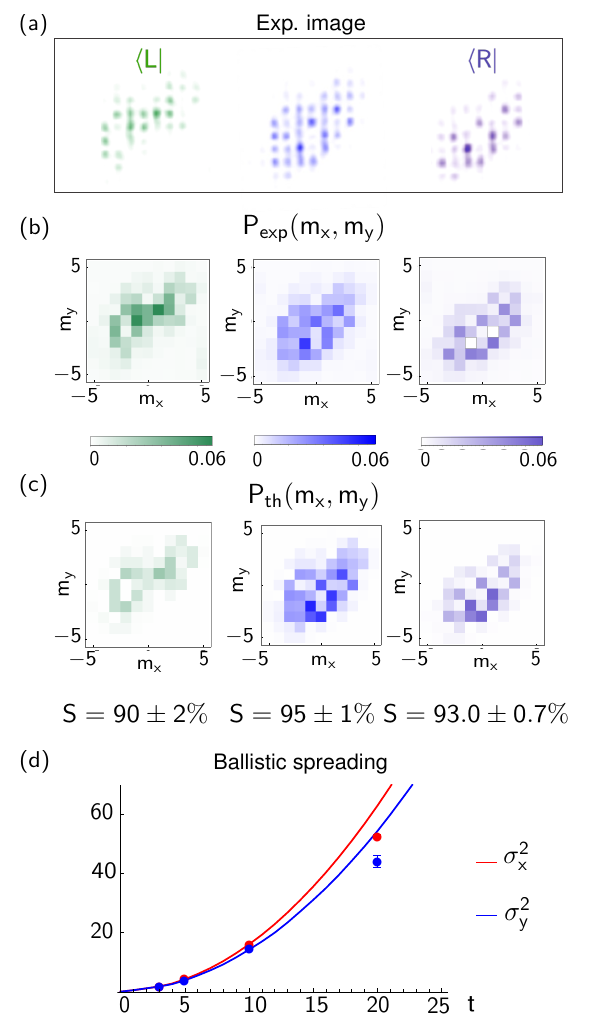}
\caption{\textbf{Resolving the totality of the modes.} A $g$-plate with a smaller spatial period $\Lambda_g \ll \Lambda$ placed before the Fourier lens allows us to resolve separately light with orthogonal circular polarizations. 
(a)~Experimental images, 
(b)~experimental reconstructions $P_{\text{exp}}(m_x,m_y)$ and (c)~theoretical predictions $P_{\text{th}}(m_x,m_y)$ of the output distribution and its projections on $\ket{L}$ and on $\ket{R}$. A localized $\ket{R}$-polarized input after 5 steps is considered.
(d)~Variance of the output distribution along $x$ and $y$. The experimental points (dots) correctly reproduce the expected ballistic behavior (solid lines) extracted numerically.}
\label{fig:variances1}
\end{figure}

Simple propagation through a lens does not allow accessing all output modes. At each spot we are not discriminating light polarization, and intensities are given by the incoherent sum of both left and right states. However, we can use a $g$-plate \cite{DErrico2020}, that is a LCMS with a linear variation of optic-axis angle, having a spatial period ${\Lambda_g=250\,\mu\text{m} \ll \Lambda}$ before the lens. In this way, left (right) polarized modes get a large momentum kick $\Delta k_g \gg \Delta k_\perp$ in the positive (negative) direction, so that they are imaged at different positions on the camera sensor. Figure~\ref{fig:variances1}(a-c) shows the projections on $\ket{L}$ and $\ket{R}$ of the probability distribution for $5$ steps and a localized $\ket{R}\text{-}\text{polarized}$ input. Since the $g$-plate is partially tuned, a fraction of the beam does not get any polarization-dependent large kick, as such in the central part of the camera we obtain the total intensity distribution. By taking similar images after mapping circular polarization states to horizontal/vertical and diagonal/antidiagonal ones (through suitably oriented waveplates), a complete analysis of the polarization of each mode could be carried out.

Contrary to the classical case, the output mode distribution depends on the input coin state. This is a consequence of the interference among different paths, which intrinsically distinguishes the QW from its classic counterpart. Nevertheless, the quantum process always presents ballistic features~\cite{Tang2018}. 
Figure~\ref{fig:variances1}(d) shows the variance over time of the output probability distributions for our QW protocol, both in the $x$ and $y$ direction. We report the measured $\sigma_{x}^2$ and $\sigma_{y}^2$ for $3$, $5$, $10$, and $20$ time steps for a localized $\ket{R}$-polarized input. The ballistic trend ${\sigma^2\propto t^2}$ is well captured in our experiments. Deviations at 20 time steps are probably due to a larger fraction of the field that remains close to the central mode (see Fig.\ \ref{fig:exp_data}(c-d), bottom row). Variance plots relative to different input polarizations are provided in the Supplementary Material.

\begin{figure*}[t]
    \includegraphics[width=0.95\linewidth]{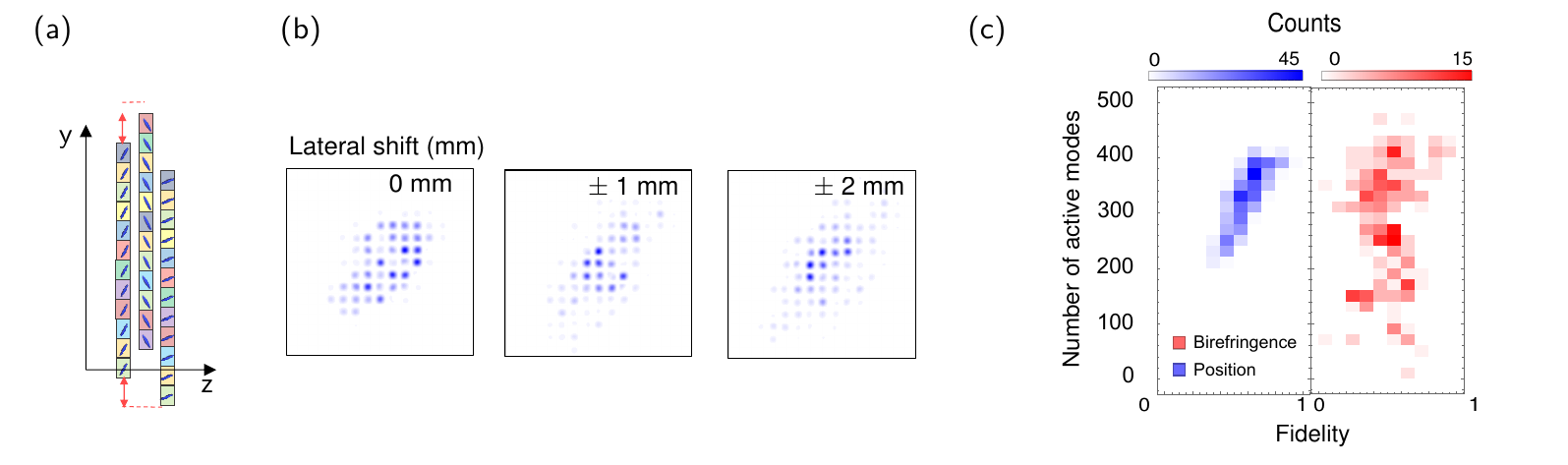}
    \caption{
    \textbf{Unitary maps obtained by reconfiguring a sequence of three plates.}
    (a)~LCMSs' optical birefringence parameters ${\delta_i \in [0,2\pi)}$, represented as the tilt of the LC molecules with respect to the propagation axis $z$, can be electrically tuned. Moreover, their lateral relative position can also be adjusted, both in the $x$ and $y$ direction (red arrows).
    (b)~When shifting the plates, the overall transformation is still a unitary circuit coupling transverse wavevector modes. The three panels show the output intensity distribution computed numerically for a $\ket{R}$-polarized input state when the LCMSs designed to implement the 5-step QW are not shifted, and when the second and the third are laterally shifted in opposite directions along both $x$ and $y$ of $\pm 1 \text{ mm}$ and $\pm 2 \text{ mm}$, respectively.
    (c)~$500$ unitary maps $\tilde{U}$ are numerically generated by randomly varying the birefringence parameters, with ${\delta_i \in [0,2\pi)}$ (red), and 500 more by randomly varying the relative position in a range $ \leq 0.15$~mm (blue) of the three LCMSs implementing the $20$-step QW. Through the histogram, the distribution of generated unitaries can be investigated, both in terms of the number of activated output modes and of the fidelity $F{(\tilde{U})}$ with respect to the reference QW process.}   
    \label{fig:fidelity}
\end{figure*}

In the experiments described above, LCMS patterns have been computed to yield the transformation associated with a target QW. To reproduce the correct map, they must be stacked carefully matching their transverse modulations to make Eq.~\eqref{eqn:solve} valid in each point. Moreover, the applied voltages must be adjusted so that they work as half-wave and quarter-wave plates. In its current implementation, the platform cannot be reprogrammed: if the target QW changes, a new set of three plates should be fabricated with the correct pattern of optic axes. However, when changing the plates' birefringence and relative positions (see Fig.~\ref{fig:fidelity}(a)), the overall transformation remains a unitary mode coupler for the transverse modes defined in Eq.~\eqref{opticalmodes}. This result is not trivial if one considers that these modes are a discrete subset in a continuum of modes associated with the transverse wavevector, which is a 2D continuous variable. In Fig.~\ref{fig:fidelity}(b), we compare the output intensity distributions computed numerically when adding lateral shifts to the LCMSs designed for a 5-step QW. Importantly, the output field corresponds to a well-defined grid of Gaussian modes in all cases.

To provide a quantitative analysis of the properties of achievable transformations, we computed the number of times a lattice mode $\ket{m_x,m_y}$ is activated when varying some of the adjustable parameters. In particular, an output state $\ket{m_x,m_y}$ is considered to be active when its intensity is above the threshold value $1/d^2$, where ${d=2 t + 1}$ and $t$ is the number of QW steps. This value corresponds to the intensity of a flat probability distribution of $d^2$ lattice modes. The LCMSs implementing the 20-step QW are chosen for reference. We also computed the fidelity of each obtained transformation $\tilde{U}$ with respect to the reference unitary process, i.e., $F(\tilde{U})=\frac{1}{2} \abs{\text{Tr}(\tilde{U}^{\dagger} U_{20})}$.
The histogram distribution of the number of active modes and the fidelity in the two configurations, when either the plates' birefringence or relative displacement is adjusted, is plotted in Fig.~\ref{fig:fidelity}(c). The latter shows that changing the plates' birefringence can significantly alter the connectivity of the circuit and produce a unitary process significantly different from the starting one, eventually approaching the identity transformation when all values of $\delta$ are close to $2\pi$. On the other hand, adjusting the plates' relative displacements (while keeping the retardations fixed) has a less pronounced impact on these aspects.
A more detailed analysis of the properties of the achievable transformations goes beyond the scope of the present work and will be investigated in the near future.

\section{Discussion}

We realized a compact photonic circuit that implements unitary transformations associated with 2D QWs on transverse modes of structured light, propagating in free space. Compressing multiple-step QW dynamics into a limited number of spin-orbit devices leads to greater complexity in their optic-axis patterns while keeping the size of the setup the same. 

The complexity of the explorable evolutions is currently limited by our fabrication routine, but this can be overcome in the future by optimizing specific stages of the procedure, or by choosing a different type of spin-orbit devices, like dielectric metasurfaces featuring sub-wavelength resolution~\cite{Yu2014meta}.

Our platform is versatile, scalable, and couples optical modes of free space propagation, with partial reconfigurability given by the tunable birefringence parameter and the relative displacements of the plates. 
This reconfigurability might be further amplified by replacing our metasurfaces with LC displays with locally tunable birefringence. However, these typically operate in reflection mode, while our platform works in transmission, with more than 85$\%$ of the input light transmitted by each device. 

The unitary transformations we have presented are not arbitrary and are inherently characterized by translation invariance. The roadmap to leverage this approach to realize more general transformations, possibly universal, necessarily requires considering diffraction between consecutive LCMSs to break the translation symmetry, using for instance concepts already demonstrated for multi-plane light converters~\cite{Morizur20100,Carpenter2019mplc}.

The limited amount of losses will allow employing these circuits to explore multi-photon evolutions, by leveraging also novel detection systems like SPAD arrays~\cite{Makowski}, single photon cameras with high temporal resolution~\cite{Nomerotski_2023_A,Zia2023p1}, or ultra-sensitive cameras based on superconducting nanowires detectors~\cite{Oripov2023}.

\appendix

\begin{figure*}[t]
    \includegraphics[width=\linewidth]{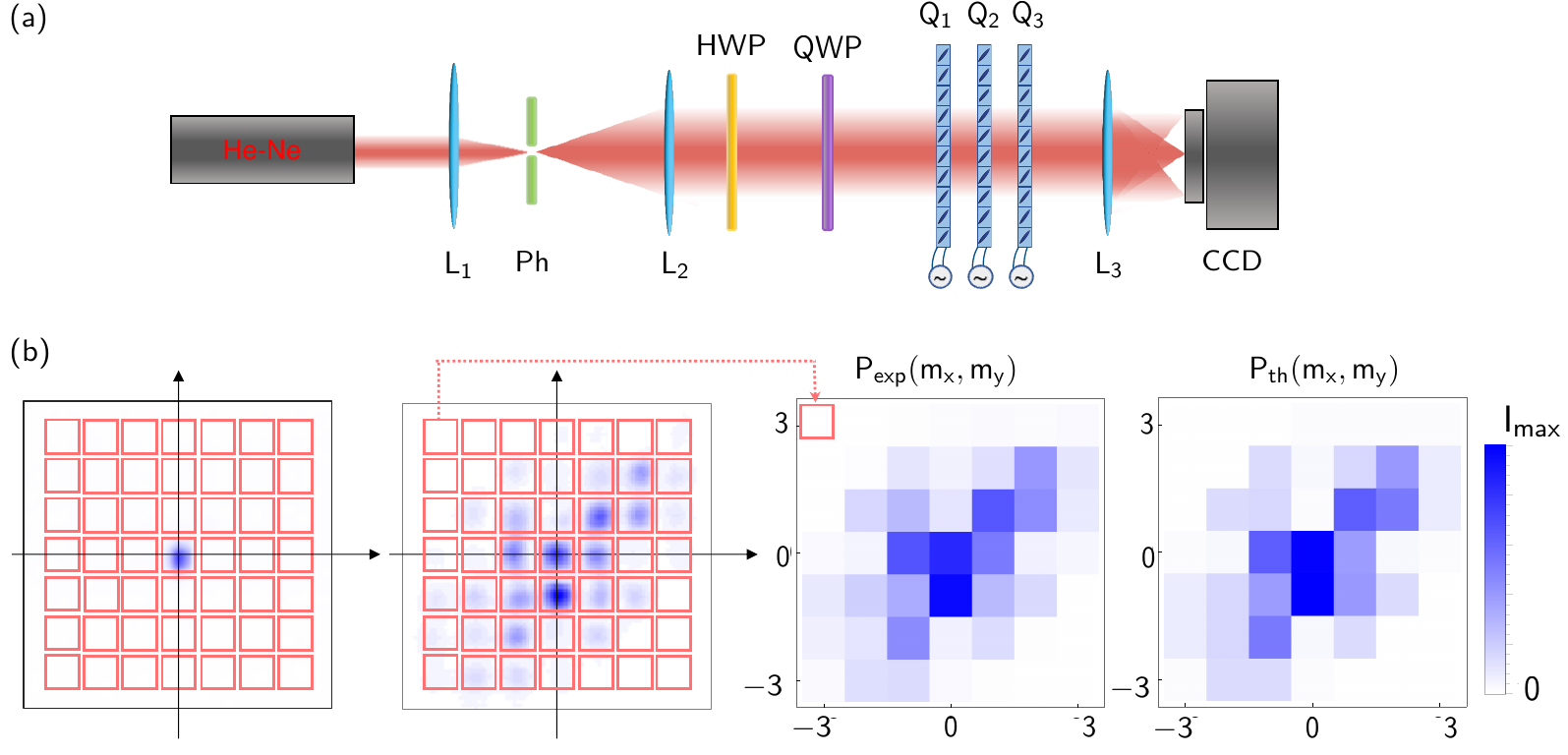}
    \caption{\textbf{Experimental implementation.} 
    (a)~Experimental setup to engineer QW dynamics. The entire evolution is compressed within only three LCMSs. 
    (b)~Reconstruction of the probability distribution $P_{\text{exp}}(m_x,m_y)$ from the experimental image. After the central ${\ket{m_x,m_y}=\ket{0,0}}$ spot has been determined, the probability of each site is computed as the normalized integrated intensity within the corresponding light spot.}
    \label{fig:setup}
\end{figure*}

\section{Experimental Setup}\label{sec:methods1}
\noindent

Our experiments are realized with the setup sketched in Fig.~\ref{fig:setup}(a).~A He–Ne laser beam (wavelength ${\lambda= 633 }$~nm) passes through a telescope system, consisting of two aspheric lenses $L_1$ and $L_2$ (with focal lengths ${f_1 =5}$~cm and ${f_2 =30}$~cm) and a $25$~$\mu$m-pinhole (Ph). The latter operates as a spatial filter. As discussed in the text, a convenient choice for the beam waist is ${w_0 \simeq \Lambda=5}$~mm. A combination of a half-wave plate (HWP) and a quarter-wave plate (QWP) sets the desired coin-polarization input state. The beam then propagates through the three LCMSs implementing the full dynamics. These are mounted in practical mounts allowing us to adjust their transverse displacement with micrometric precision, both in the $x$ and $y$ direction. This is needed for an accurate alignment of the plates, which makes Eq.~\eqref{eqn:solve} valid at each point. At the exit of the last metasurface, we set a lens $L_3$ (with focal length $f_3 = 50$~cm), Fourier-transforming light momenta into positions on the CCD camera placed in the focal plane.

\section{Analytical solutions for the metasurfaces' patterns}\label{sec:methods2}

Here we provide the analytical solutions to Eq.~\eqref{process}.
In our optical encoding, a single-step QW operator $U_0$ can be conveniently visualized as a space-dependent polarization transformation~\cite{DiColandrea2023}:
\begin{equation}
U_0=\iint \text{d}x\text{d}y\,\,\mathcal{U}_0(x,y)\ketbra{x,y},
\end{equation}
with $\mathcal{U}_0$ an SU(2) operator: 
\begin{equation}
\mathcal{U}_0(x,y)=\cos E\left(x,y\right)\sigma_0-i\sin E\left(x,y\right)\textbf{n}\left( x,y\right)\cdot\bm{\sigma},
\end{equation}
where ${E}$ is a real parameter defined in the range ${[0,\pi]}$ and ${\textbf{n}=\left(n_1,n_2,n_3\right)}$ is a unit vector. At each transverse position, the operator $U_0$  implements a polarization rotation ${\mathcal{U}_0(x,y)}$ of an angle $2E$ around the axis $\textbf{n}$.

In our experiment, the target operator is a $t$-step QW, obtained as ${U=U_0^t}$. Accordingly, the space-dependent target operator ${\mathcal{U}=\mathcal{U}_0^t}$ can be decomposed as
\begin{equation}
\mathcal{U}(x,y)=\cos\xi\left(x,y\right)\sigma_0-i\sin\xi\left(x,y\right)\textbf{n}\left( x,y\right)\cdot\bm{\sigma},
\end{equation}
where ${\xi(x,y)=E(x,y)t}$.

 The optical sequence of three LCMSs, ${\mathcal{L}=Q_{\pi/2}(\theta_3)Q_{\pi}(\theta_2)Q_{\pi/2}(\theta_1)}$, is analogously decomposed as

\begin{equation}
\begin{aligned}
    \mathcal{L}(x,y) &= \ell_0(x,y)\sigma_0 \\
    &\quad - i\left(\ell_1(x,y)\sigma_1 + \ell_2(x,y)\sigma_2 + \ell_3(x,y)\sigma_3\right),
\end{aligned}
\end{equation}
where 
\begin{equation}
\begin{split}
\ell_0 &= -\cos\alpha\cos\beta,\\
\ell_1 &= -\sin\beta\sin\gamma,\\
\ell_2 &= \sin\beta\cos\gamma,\\
\ell_3 &= \sin\alpha\cos\beta,
\end{split}
\end{equation}
with ${\alpha=\theta_1-\theta_3}$, ${\beta=\theta_1-2\theta_2+\theta_3}$, and ${\gamma=\theta_1+\theta_3}$. The dependence on $(x,y)$ is omitted for ease of notation. Imposing ${\mathcal{U}=\mathcal{L}}$ at each transverse position yields
\begin{subequations}
\begin{equation}
\cos\alpha\cos\beta=-\cos\xi,
\label{eqn:eqn1}
\end{equation}
\begin{equation}
\sin\beta\sin\gamma=-\sin\xi\sin\theta\cos\phi,
\label{eqn:eqn2}
\end{equation}
\begin{equation}
\sin\beta\cos\gamma=\sin\xi\sin\theta\sin\phi,
\label{eqn:eqn3}
\end{equation}
\begin{equation}
\sin\alpha\cos\beta=\sin\xi\cos\theta,
\label{eqn:eqn4}
\end{equation}
\end{subequations}
where we used the spherical parametrization of the vector $\textbf{n}$: ${n_1=\sin\theta\cos\phi}$, ${n_2=\sin\theta\sin\phi}$, and ${n_3=\cos\theta}$.

\begin{figure}[h]
\centering
\includegraphics[width=\linewidth]{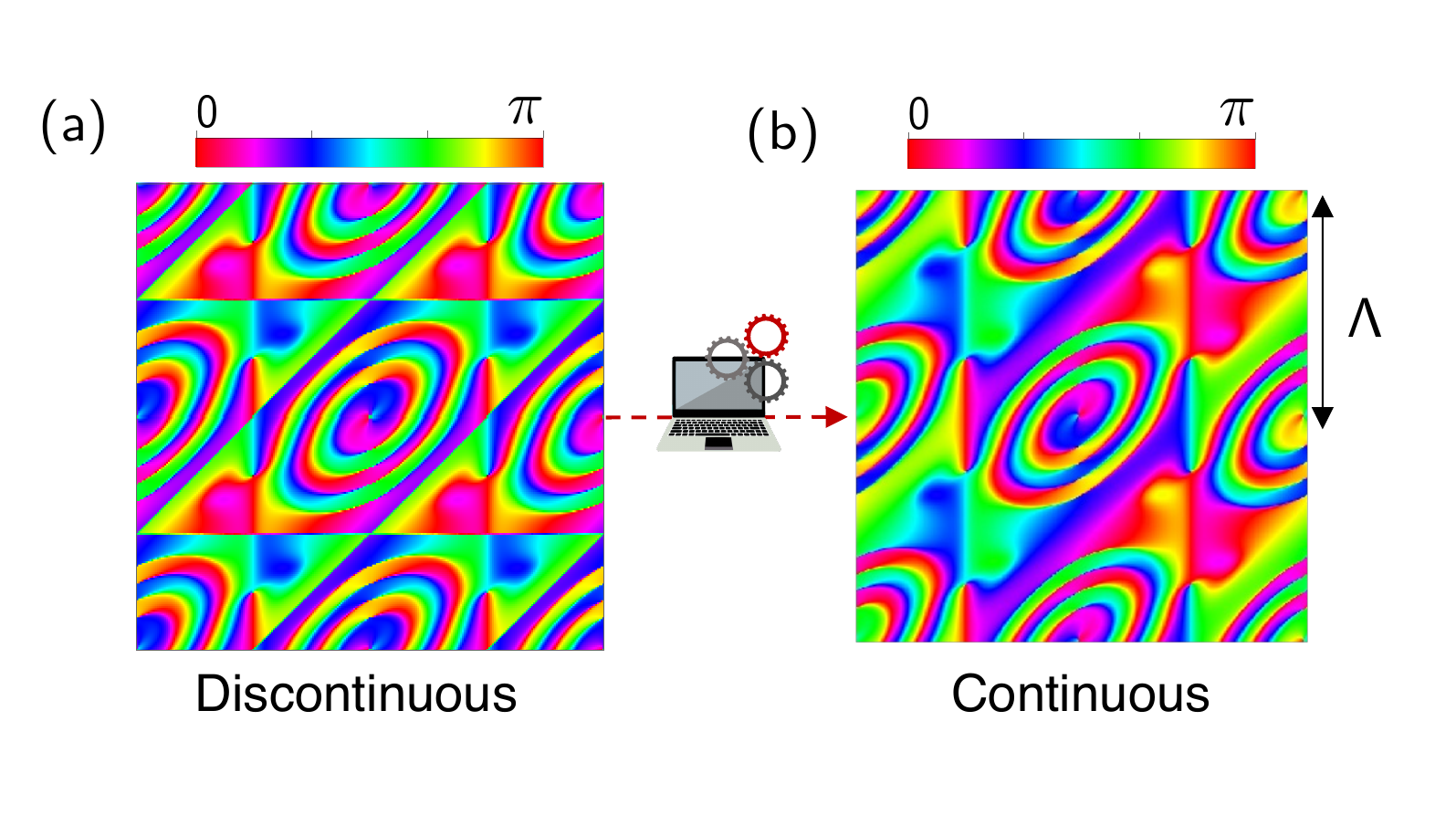}
\caption{\textbf{Optimization routine.} 
(a)~Single analytical solution for the first LCMS implementing the QW evolution ${U_0^{10}(\pi/2)}$. Discontinuities across extended lines are visible.  
(b)~The optimization routine is executed to output a continuous modulation for the sets of three LCMSs, only embedding isolated vortices.}
\label{fig:jumpingsol}
\end{figure}

From Eq.~\eqref{eqn:eqn2} and Eq.~\eqref{eqn:eqn3} it follows:
\begin{equation}
\gamma=\phi-\frac{\pi}{2}.
\end{equation}
Two sets of solutions are found from Eq.~\eqref{eqn:eqn1} and Eq.~\eqref{eqn:eqn4}:
\begin{equation}
\begin{cases}
\alpha_1=\text{atan2}\left(-\sin\xi\cos\theta,\cos\xi\right)\\
\beta_1=\text{atan2}\left(\sin\xi\sin\theta,-\sqrt{1-\sin^2\xi\sin^2\theta}\right)\\
\gamma_1=\phi-\pi/2
\end{cases}
\end{equation}

\begin{equation}
\begin{cases}
\alpha_2=\pi+\alpha_1\\
\beta_2=\pi-\beta_1\\
\gamma_2=\gamma_1
\end{cases}
\end{equation}
where atan2(${x,y}$) is the two-argument arctangent function, which distinguishes between diametrically opposite directions. From the expressions for $\alpha$, $\beta$, and $\gamma$, the modulations for the LCMS's optic-axis patterns $\theta_1$, $\theta_2$, and $\theta_3$ can be finally extracted.
As discussed in the text, a modulation extracted from a single set of solutions typically features discontinuities and sudden jumps. As a representative example, in Fig.~\ref{fig:jumpingsol}(a), we plot the pattern of the first LCMS emerging from a single set of solutions for ${t=10}$. The optimization routine described in the text (see Fig.~\ref{fig:search}(a)) is executed to match the two sets of solutions and eventually remove the discontinuities (see Fig.~\ref{fig:jumpingsol}(b)).   

\section{Reconstruction of the probability distribution}\label{sec:methods3}
To reconstruct the QW probability distribution from the experimental image, the coordinates of the lattice site $\ket{m_x,m_y}=\ket{0,0}$ have to be determined. To do so, we set the birefringence parameter of the three LCMSs to ${\delta_1=\delta_2=\delta_3=2\pi}$. In this configuration, the metasurfaces act as the identity operator at each point $(x,y)$, and only the central input mode is transmitted.  
Starting from the coordinates of the corresponding spot in the focal plane of $L_3$, we build an array of equally spaced square regions on the image, each associated with an output mode.
Then, we set ${\delta_1=\delta_3=\pi/2}$ and ${\delta_2=\pi}$, so that the plates implement the desired complex polarization transformation which simulates the target dynamics $\mathcal{U}(x,y)$. By integrating light intensity within each square, which provides a good estimate of the amount of input light coupled to each output mode, and normalizing it to the total intensity, we eventually reconstruct the experimental probability distribution $P_{\text{exp}}(m_x,m_y)$. This procedure is depicted in Fig.~\ref{fig:setup}(b).
The agreement between theory and experimental results is quantified in terms of the similarity estimator $S$ (cf.~Eq.~\eqref{similarity}).

\bibliography{main}

\subsection*{Acknowledgements}
We acknowledge Alexandre Dauphin and Alioscia Hamma for fruitful discussions. This work was supported by PNRR MUR project PE0000023-NQSTI.
\subsection*{Author contributions}
FC conceived the idea, FDC performed numerical simulations and extracted the patterns of the LCMSs. MGA and FDC performed the experiment and the data analysis. LM and FC supervised the project. All authors discussed the results and contributed to the writing of the manuscript.
\subsection*{Disclosures}
The authors declare that they have no competing interests.
\subsection*{Data and materials availability}
All data are available in the main text or the supplementary materials.

\clearpage
\onecolumngrid
\renewcommand{\figurename}{\textbf{Figure}}
\setcounter{figure}{0} \renewcommand{\thefigure}{\textbf{S{\arabic{figure}}}}
\setcounter{table}{0} \renewcommand{\thetable}{S\arabic{table}}
\setcounter{section}{0} \renewcommand{\thesection}{S\arabic{section}}
\setcounter{equation}{0} \renewcommand{\theequation}{S\arabic{equation}}
\onecolumngrid

\begin{center}
{\Large Supplementary Material for: \\ Large-scale free-space photonic circuits in two dimensions}
\end{center}
\vspace{1 EM}

\section*{Supplementary Data}
\begin{figure*}[h]
    \includegraphics[width=\linewidth]{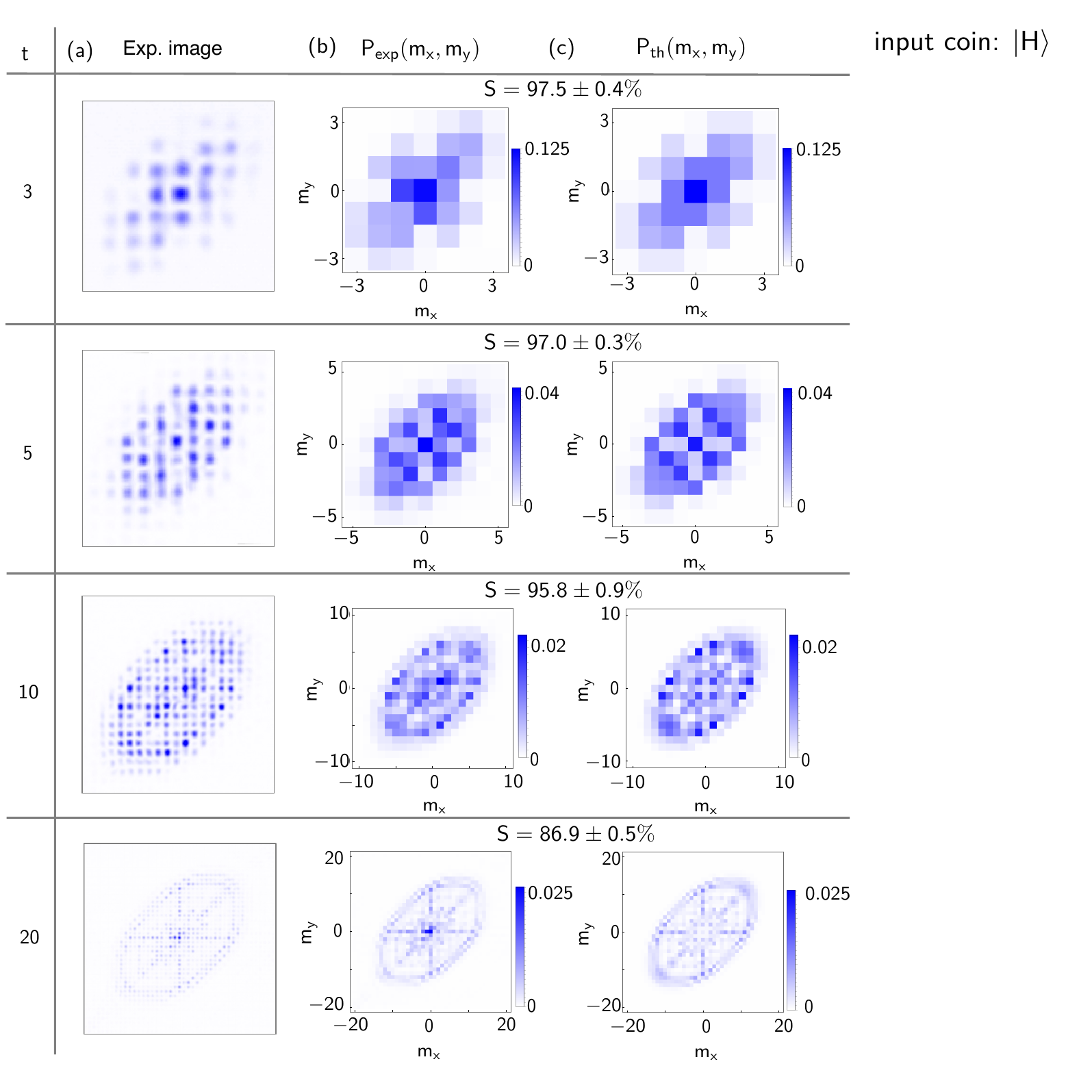}
    \caption{\textbf{2D QWs via spin-orbit photonics.} (a)~Experimental images obtained for a $\ket{H}$-polarized input state, from which the walker probability distribution $P_{\text{exp}}(m_x,m_y)$ is extracted (b), and compared with the theoretical prediction $P_{\text{th}}(m_x,m_y)$ (c). For each realization, we report the value of the similarity, computed as the average of four independent measurements. The rows refer to $3$, $5$, $10$, and $20$ time steps ($t$), respectively.}
    \label{fig:supplementary1H}
\end{figure*}

\begin{figure*}[h]
    \includegraphics[width=\linewidth]{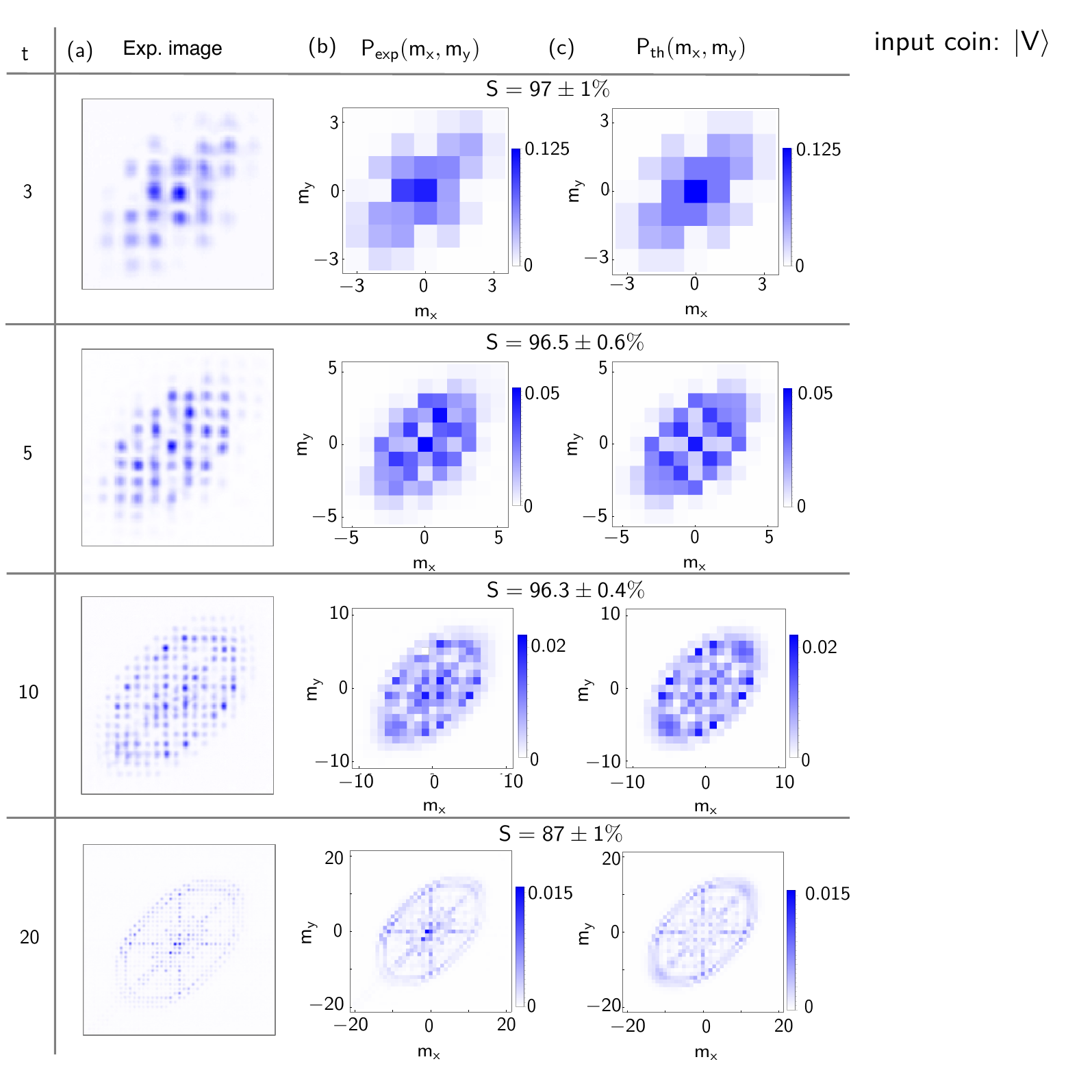}
    \caption{\textbf{2D QWs via spin-orbit photonics.} (a)~Experimental images obtained for a $\ket{V}$-polarized input state, from which the walker probability distribution $P_{\text{exp}}(m_x,m_y)$ is extracted (b), and compared with the theoretical prediction $P_{\text{th}}(m_x,m_y)$ (c). For each realization, we report the value of the similarity, computed as the average of four independent measurements. The rows refer to $3$, $5$, $10$, and $20$ time steps ($t$), respectively.}
    \label{fig:supplementary1V}
\end{figure*}

\begin{figure*}[h]
    \includegraphics[width=\linewidth]{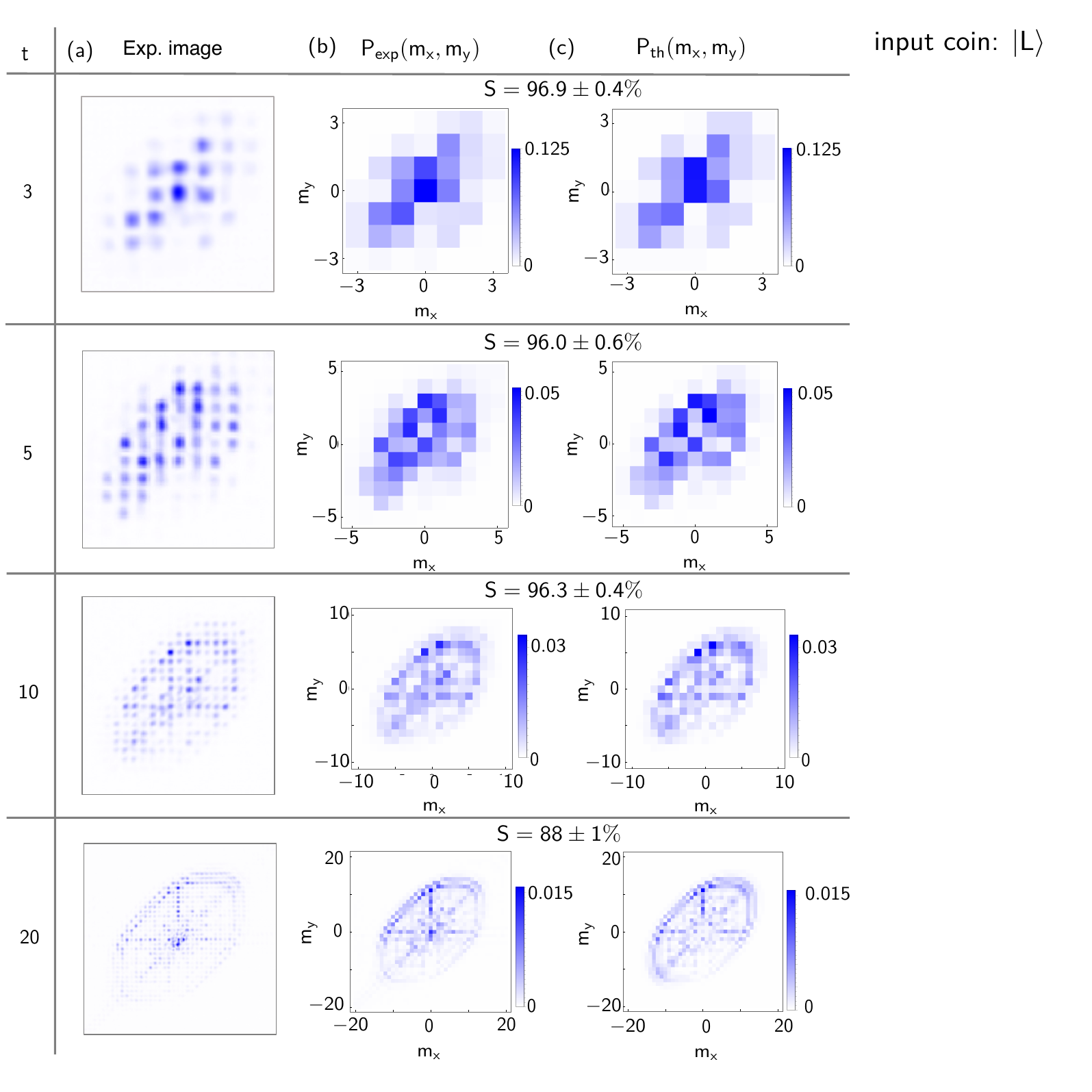}
    \caption{\textbf{2D QWs via spin-orbit photonics.} (a)~Experimental images obtained for a $\ket{L}$-polarized input state, from which the walker probability distribution $P_{\text{exp}}(m_x,m_y)$ is extracted (b), and compared with the theoretical prediction $P_{\text{th}}(m_x,m_y)$ (c). For each realization, we report the value of the similarity, computed as the average of four independent measurements. The rows refer to $3$, $5$, $10$, and $20$ time steps ($t$), respectively.}
    \label{fig:supplementary1L}
\end{figure*}

\begin{figure*}[h]
    \includegraphics[width=\linewidth]{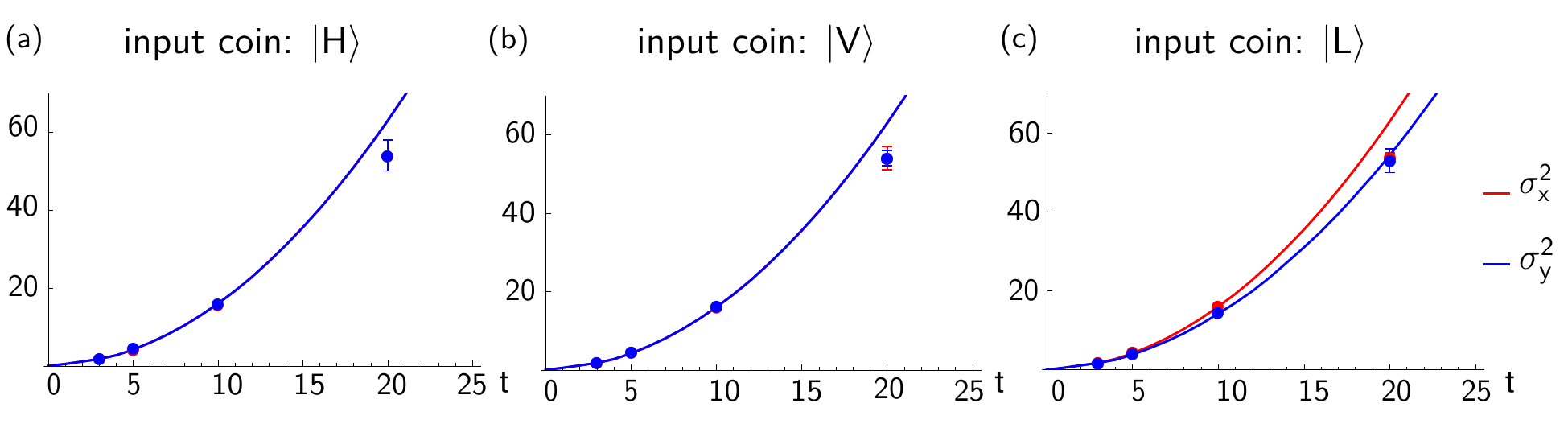}
    \caption{\textbf{Variance of 2D QW distributions.} Variance of the output distribution along $x$ and $y$, $\sigma_x^2$ and $\sigma_y^2$, for different input states: (a) $\ket{H}$, (b) $\ket{V}$, (c) $\ket{L}$. The experimental points (dots) correctly reproduce the expected ballistic behavior (solid lines) extracted numerically.}
    \label{fig:supplementary2}
\end{figure*}

We provide experimental results for different input coin-polarization states. Figures~\ref{fig:supplementary1H}-\ref{fig:supplementary1V}-\ref{fig:supplementary1L} show the probability distributions obtained for a $\ket{H}$, $\ket{V}$, and $\ket{L}$ polarization input state (cf.~Fig.
~\textcolor{blue}{3}), respectively. Figure~\ref{fig:supplementary2} shows the QW ballistic spreading for the same polarization inputs (cf.~Fig.
~\textcolor{blue}{4}(d)). 

\begin{figure*}[h]
    \includegraphics[width=\linewidth]{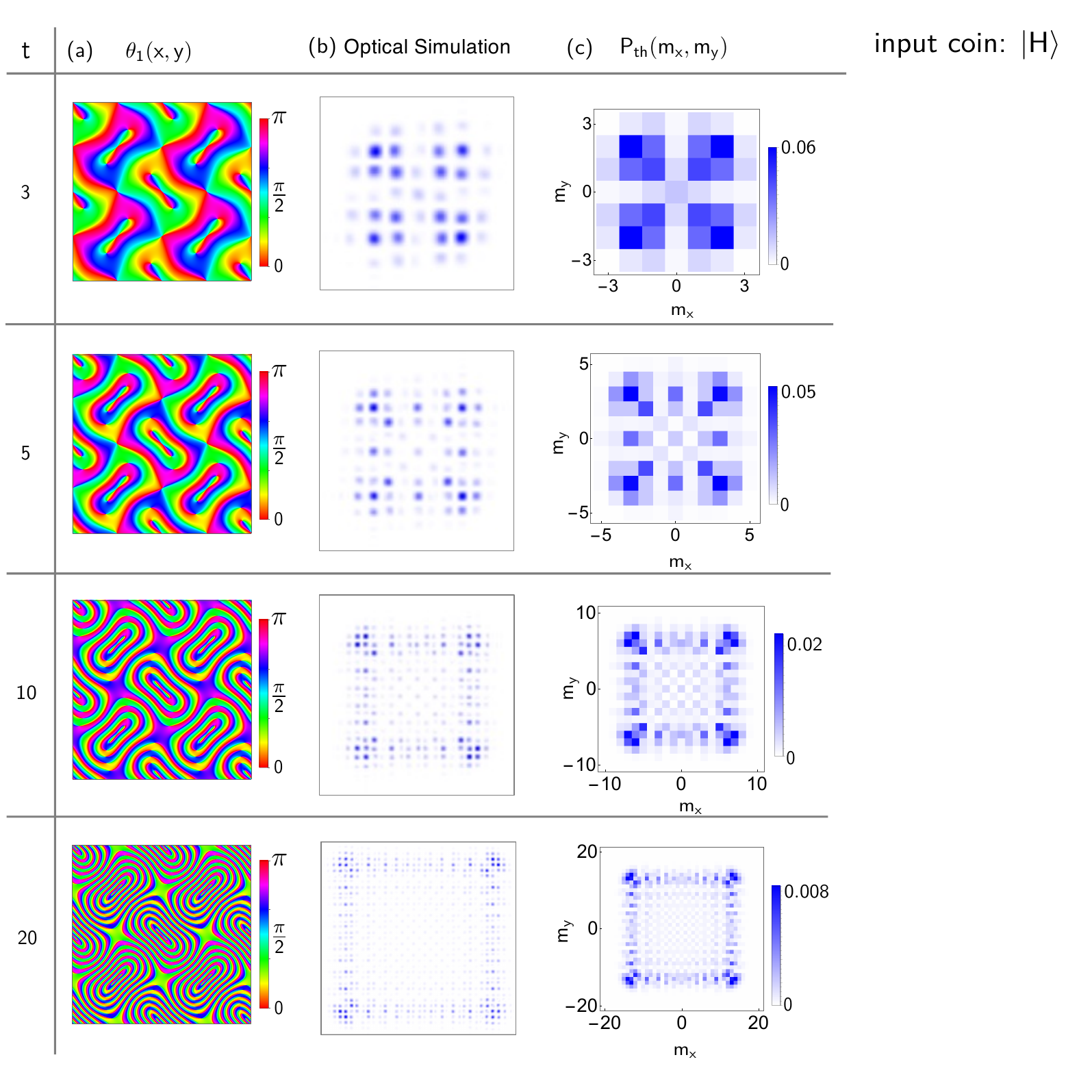}
    \caption{\textbf{Split-step 2D QW with LCMSs.} (a)~Computed optic-axis modulation of the first metasurface ($\theta_1(x,y)$) for the simulation of a split-step 2D QW at different time steps ($t$). (b)~Simulation of the output field intensity for a $\ket{H}$-polarized input, compared with the theoretical prediction (c) $P_{\text{th}}(m_x,m_y)$.}
    \label{fig:supplementarysplit}
\end{figure*}

To demonstrate the versatility of our method, we compute the optic-axis patterns of the 3 LCMSs that would be needed for the simulation of 3, 5, 10, and 20 time steps of an alternative split-step QW protocol:
${U_{ss}=T_y(\delta=\pi/2) W T_x(\delta=\pi/2) W}$, where $T_x$, $T_y$ and $W$ are the lattice and coin operators introduced in the manuscript. We also simulate the output field intensity for a localized $\ket{H}$-polarized input state and compare it with the theoretical QW probability distribution. The results are reported in Fig.~\ref{fig:supplementarysplit}.

\end{document}